\documentstyle[12pt]{article}
\begin{document}
\begin{center}
\baselineskip 0.70cm
\vskip 0.4in
{\Large \bf Ferromagnetism and Colossal Magnetoresistance  
from the 
Coexistence of Comparable Charge and Spin Density Orders}
\vskip 0.3in

Georgios VARELOGIANNIS \footnote{varelogi@mpipks-dresden.mpg.de}

{\it Max-Planck-Institut f\"ur Physik Komplexer Systeme\\ 
N\"othnitzer Str. 38,                               
01187 Dresden, Germany} 
\vskip 0.15in

\vskip 0.6in
\begin{abstract}
\baselineskip 0.80cm
We report a complete multicomponent mean-field-theory 
for the coexistence and competition of
charge ordering (CO),
antiferromagnetic (AFM)
and ferromagnetic (FM)
spin ordering in the presence of a uniform
magnetic field. 
Doping the
AFM or CO state
always generates a ferromagnetic
component. Itinerant FM, AFM and CO,
necessarily coexist and compete 
in a particle-hole asymmetric system.
Melting
of large AFM-CO orders by small magnetic fields
and the related phenomenon of
Colossal Magnetoresistance (CMR)
may arise
whenever 
the CO and AFM order
parameters have similar magnitude and momentum structure.
Hole doping favors FM metallic states and CMR while
electron doping favors AFM-CO states in
agreement with the phase diagram of perovskite manganites.
\end{abstract}
\end{center}
\newpage
\baselineskip 0.75cm

The pervovskite manganites $(La,Pr)_{1-x}(Ca,Sr,Ba)_xMnO_3$, 
in the doping region $x\approx 0.2-0.4$
exhibit a transition 
to a ferromagnetic (FM) ground state which is accompanied by a 
large drop of the resistivity \cite{Ramirez}.
This transition can be tuned by 
the application of a magnetic field producing
negative
``Colossal Mangetoresistance'' (CMR) \cite{Jin}.
Despite the intense experimental and theoretical efforts,
many fundamental issues are still under debate including
the physical origin of the CMR phenomenon.
Ferromagnetism in these 
materials is usually attributed to the double exchange
mechanism \cite{DE,DE1}, in which the    
lattice 
degrees of freedom \cite{MLS,millis} might also be 
involved. 
However, the
CMR phenomenon could be more general
since it has also been observed in pyrochlore manganites
\cite{pyroexp}, where double exchange and 
Jahn-Teller effects on the transport can be safely
excluded \cite{noDE,PBL}. 

One of the most puzzling aspects        
of 
perovskite manganites is that
the hole doped $(x< 0.5)$ and the electron doped $(x>0.5)$
compounds behave very differently. 
In the intermediate
doping region $x\approx 0.5$ there is a kind of         
boundary between the hole doped regime where the
metallic ferromagnetic phases and CMR take place and the 
electron doped regime where essentially there are phases
of coexisting charge and spin ordering.
Understanding the physics in this intermediate region
$x\approx 0.5 \pm \varepsilon$ appears crucial,                 
and much of the 
recent experimental activity has focused on it
\cite{x05a,x05b,x05c,x05d,x05e,x05g,x05f,Liaro}
reporting some additional puzzling facts.
The coexistence 
of the AFM charge ordered (AFM-CO) state with 
FM metallic state has been established
\cite{x05f,x05d,x05e,x05g}.
Apparently a small part of the carriers remains metallic
in the AFM-CO regime,
and has been reported that even in the hole doped regime
the carriers are   
separated into a part that is metallic
and a part that is still charge ordered \cite{x05f,Fath,Balagurov}.
Microscopic theoretical models would also support a
spatial separation of FM and AFM-CO phases \cite{Dagotto}.
Even more puzzling is the fact that
the AFM-CO state near the half-filling
boundary
can be melt by 
the application of magnetic fields of few Teslas despite the
fact that the CO gap is very large ($\approx 0.5 eV$) and
would correspond to several
hundreds of Teslas \cite{x05f,x05b,x05a}.  
The melting of the AFM-CO state appears to
progress through the increase of the number of carriers
which are FM metallic \cite{x05f}.

The above
experimental findings suggest that CO, AFM and FM coexist and
compete and a general mean-field theory is therefore necessary
in which all these order
parameters are considered self-consistently on
the same footing. Such theory is
reported for the first time in this
Letter and surprisingly, not only provides a natural understanding
of the above  
puzzling behavior in the intermediate doping region
of perovskite manganites
but also provides unexpected fundamental insight in the 
underlying physics of CMR and itinerant FM.
Other mean-field theories have considered the above
orders but only one by one and therefore cannot account
for their coexistence to which our original results are due.

We study the general mean-field hamiltonian
describing the coexistence of
CO, AFM and FM orders in the
presence of a uniform magnetic field 
$$
H = \sum_{\bf{k},\alpha}\xi_{\bf{k}\alpha}
c^{\dagger}_{\bf{k}\alpha}c_{\bf{k}\alpha}-
\sum_{\bf{k},\alpha,\beta}\delta_{\alpha\beta}W_{\bf k}\biggl(
c^{\dagger}_{\bf{k}\alpha}c_{\bf{k}+\bf{Q}\beta} + HC\biggr)
$$
$$
-\sum_{\bf{k},\alpha,\beta}({\bf \sigma}\cdot {\bf n})_{\alpha\beta}
M_{\bf k}\biggl(
c^{\dagger}_{\bf{k}\alpha}c_{\bf{k}+\bf{Q}\beta} + HC\biggr)
$$
$$
-
\sum_{\bf{k},\alpha,\beta}
({\bf \sigma}\cdot {\bf n})_{\alpha\beta}
\biggl(F_{\bf k}+\mu_B H \biggr)
\biggl(
c^{\dagger}_{\bf{k}\alpha} 
c_{\bf{k}\beta}
+HC
\biggr)
\eqno(1)
$$
where $\alpha,\beta$ are spin indices,
$W_{\bf k}$, $M_{\bf k}$ and $F_{\bf k}$
are the CO, AFM and FM order parameters
respectively, ${\bf n}$ 
the polarizations of the AFM and FM orders considered here parallel
without influence on the generality of the results,
$\xi_{\bf{k}}$ the electronic dispersion and $\mu_B H$ the Zeeman
contribution of the applied magnetic field.
The above hamiltonian accounts for
the physics resulting from the coexistence of the
AFM, CO and FM orders whatever the exact microscopic mechanism 
responsible for these orderings
exactly as the BCS hamiltonian
accounts for the physics related to the supercocnducting ordering
irrespective of the exact pairing mechanism.
The general conclusions of our study apply 
to any itinerant
system in which the above orders are present 
and therefore to manganites as well.

To study all order phenomena on the same footing,
we must work in a multicomponent spinor space     
\cite{multi}.
We use an 
eight component spinor formalism with a basis
defined by the following tensor products:
$
\widehat{\tau}=\widehat{\sigma}\otimes\bigl(\widehat{I}\otimes 
\widehat{I})
$, $
\widehat{\rho}=\widehat{I}\otimes\bigl(\widehat{\sigma}\otimes
\widehat{I})
$, $
\widehat{\sigma}=\widehat{I}\otimes\bigl(\widehat{I}\otimes
\widehat{\sigma})
$,
where
$\widehat{\sigma}$ are the usual Pauli matrices
and $\widehat{I}$ the identity matrix.
We define $2\gamma_{\bf k}=\xi_{\bf{k}}-\xi_{\bf{k+Q}}$
and $2\delta_{\bf k}=\xi_{\bf{k}}+\xi_{\bf{k+Q}}$.
When $\delta_{\bf k}=0$ there is
particle-hole symmetry or perfect nesting
at the wavevector ${\bf Q}$. 
With the above notations and considering all
order parameters real 
we have obtained the
one particle diagonal thermal Green's function 
corresponding 
to our hamiltonian.
It 
can be written as follows:
$$
\widehat{G}_o ( {\bf k} , i \omega_n ) =
- \lbrack
i\omega_n\widehat{\tau}_2+i
\gamma_{\bf{k}}\widehat{\tau}_1\widehat{\rho}_3
+\delta_{\bf{k}}\widehat{\tau}_2\widehat{\rho}_3
+iW_{\bf k}\widehat{\tau}_3\widehat{\rho}_3
+iM_{\bf k}\widehat{\tau}_3
\widehat{\rho}_3\widehat{\sigma}_3
+(F_{\bf k}+\mu_B H )
\widehat{\tau}_2\widehat{\rho}_3\widehat{\sigma}_3\rbrack
$$
$$
\times\biggl[
A( {\bf k} , i \omega_n )
\widehat{\tau}_2
+i2\gamma_{\bf{k}}\delta_{\bf{k}}\widehat{\tau}_1
+i[2W_{\bf{k}}\delta_{\bf{k}}+2M_{\bf k}
(F_{\bf k}+\mu_B H )]\widehat{\tau}_3
+[2W_{\bf{k}}M_{\bf{k}}+2(F_{\bf k}+\mu_B H )\delta_{\bf{k}}]
\widehat{\tau}_2\widehat{\sigma}_3
$$
$$
+i[2M_{\bf{k}}\delta_{\bf{k}}+2W_{\bf k}(F_{\bf k}+\mu_B H)]
\widehat{\tau}_3
\widehat{\sigma}_3
+i2\gamma_{\bf{k}}(F_{\bf k}+\mu_B H )\widehat{\tau}_1\widehat{\sigma}_3
\biggr]
$$
$$
\times
\biggl[ B( {\bf k} , i \omega_n )
- \Gamma ( {\bf k} , i \omega_n )\widehat{\sigma}_3\biggr]
D({\bf k} , i \omega_n )
\eqno(2)
$$
where $\omega_n=(2n+1)\pi T$ are the Matsubara frequencies for
fermions. 
To condense the formal expressions the following
functionals have been defined:
$$
A( {\bf k} , i \omega_n )=\omega^2_n+\gamma^2_{\bf{k}}+
\delta^2_{\bf{k}}+
W^2_{\bf{k}}+M^2_{\bf{k}}+(F_{\bf k}+\mu_B H)^2
\eqno(3)
$$
$$
B( {\bf k} , i \omega_n )=A^2( {\bf k} , i \omega_n )
-4\gamma^2_{\bf{k}}\delta^2_{\bf{k}}-4[W_{\bf k}\delta_{\bf{k}}+
M_{\bf{k}}(F_{\bf k}+\mu_B H)]^2
+4[W_{\bf k}M_{\bf{k}}+\delta_{\bf{k}}(F_{\bf k}+\mu_B H)]^2
$$
$$
-4[M_{\bf k}\delta_{\bf{k}}+
W_{\bf{k}}
(F_{\bf k}+\mu_B H)]^2-4\gamma^2_{\bf{k}}(F_{\bf k}+\mu_B H)^2
\eqno(4)
$$
$$
\Gamma( {\bf k} , i \omega_n )=
4A( {\bf k} , i \omega_n )[W_{\bf k}M_{\bf{k}}+\delta_{\bf{k}}
(F_{\bf k}+\mu_B H)]
-8\gamma^2_{\bf{k}}\delta_{\bf{k}}(F_{\bf k}+\mu_B H)
$$
$$
-8[W_{\bf k}\delta_{\bf{k}}+M_{\bf{k}}
(F_{\bf k}+\mu_B H)]
[M_{\bf k}\delta_{\bf{k}}+
W_{\bf{k}}(F_{\bf k}+\mu_B H)]
\eqno(5)
$$
$$
D( {\bf k} , i \omega_n )=
\biggl[
[\omega^2_n+E^2_{++} ({\bf k})]
[\omega^2_n+E^2_{+-} ({\bf k})]
[\omega^2_n+E^2_{-+} ({\bf k})]
[\omega^2_n+E^2_{--} ({\bf k})]
\biggr]^{-1}
\eqno(6)
$$
We obtain four different quasiparticle branches
$
E_{\pm\pm} ({\bf k})$  
defined as follows:
$$
E_{+\pm} ({\bf k})=
\sqrt{
\gamma^2_{\bf{k}}+(W_{\bf k} \pm M_{\bf k})^2}
+ [\delta_{\bf{k}}\pm (F_{\bf k}+\mu_B H)]
\eqno(7)
$$
$$
E_{-\pm} ({\bf k})=
\sqrt{
\gamma^2_{\bf{k}}+(W_{\bf k} \pm M_{\bf k})^2}
- [\delta_{\bf{k}}\pm (F_{\bf k}+\mu_B H)]
\eqno(8)
$$

The order parameters $W_{\bf k}$, $M_{\bf k}$ and  
$F_{\bf k}$ obey self-consistency relations (e.g.
$
W_{{\bf k}}= T \sum_{{\bf k'}}\sum_{n}
V_{{\bf k k'}}^{CO} {1\over 8}
Tr
\biggl\{
\widehat{\tau}_1\widehat{\rho}_3
\widehat{G}_o
({\bf k'}, i\omega_n )\biggr\}
$ etc.).
The requirement of self-consistency leads to a   
system of coupled equations which 
are reported here because they are necessary for the following discussion
$$
W_{{\bf k}}
=-T \sum_{{\bf k'}}\sum_{n}V_{{\bf k k'}}^{CO}
\Biggl\{
W_{\bf k'}
f( {\bf k'} , i \omega_n )
+M_{\bf k'}g( {\bf k'} , i \omega_n )
$$
$$
-(F_{\bf k'}+\mu_B H)h( {\bf k'} , i \omega_n )
-\delta_{\bf{k'}}u( {\bf k'} , i \omega_n )\Biggr\} 
D( {\bf k'} , i \omega_n )
\eqno(9)
$$
$$
M_{{\bf k}}
=-T \sum_{{\bf k'}}\sum_{n}V_{{\bf k k'}}^{AFM}
\Biggl\{
M_{\bf k'}
f( {\bf k'} , i \omega_n )
+W_{\bf k'}g( {\bf k'} , i \omega_n )
$$
$$
-(F_{\bf k'}+\mu_B H)u( {\bf k'} , i \omega_n )
-\delta_{\bf{k'}}h( {\bf k'} , i \omega_n )\Biggr\}
D( {\bf k'} , i \omega_n )
\eqno(10)
$$
$$
F_{{\bf k}}
=-T \sum_{{\bf k'}}\sum_{n}V_{{\bf k k'}}^{FM}
\Biggl\{
(F_{\bf k'}+\mu_B H)
f( {\bf k'} , i \omega_n )
+\delta_{\bf{k'}}
g( {\bf k'} , i \omega_n )-y( {\bf k'} , i \omega_n )
$$
$$
-W_{\bf k'}h( {\bf k'} , i \omega_n )
-M_{\bf k'}u( {\bf k'} , i \omega_n )\Biggr\}
D( {\bf k'} , i \omega_n )
\eqno(11)
$$
where
$$
f( {\bf k} , i \omega_n )=
A({\bf k}, i\omega_n )B({\bf k},i\omega_n)
-2\bigl[W_{\bf k}M_{\bf{k}}+\delta_{\bf{k}}
(F_{\bf k}+\mu_B H)
\eqno(12)
$$
$$
g( {\bf k} , i \omega_n )=2\bigl[W_{\bf{k}} M_{\bf k}+\delta_{\bf{k}}
(F_{\bf k}+\mu_B H)\bigr]
B( {\bf k} , i \omega_n )-A( {\bf k} , i \omega_n )
\Gamma( {\bf k} , i \omega_n )
\eqno(13)
$$
$$
h( {\bf k} , i \omega_n )=
2\bigl[M_{\bf{k}}\delta_{\bf{k}}+W_{\bf k}
(F_{\bf k}+\mu_B H)\bigr]B( {\bf k} , i \omega_n )
-2\bigl[W_{\bf k}\delta_{\bf{k}}+M_{\bf{k}}
(F_{\bf k}+\mu_B H)\bigr]\Gamma( {\bf k} , i \omega_n )
\eqno(14)
$$
$$
u( {\bf k} , i \omega_n )=
2\bigl[W_{\bf{k}}\delta_{\bf{k}}+M_{\bf k}
(F_{\bf k}+\mu_B H)\bigr]B( {\bf k} , i \omega_n )
-2\bigl[M_{\bf k}\delta_{\bf{k}}+W_{\bf{k}}
(F_{\bf k}+\mu_B H)\bigr]\Gamma( {\bf k} , i \omega_n)
\eqno(15)
$$
$$
y( {\bf k} , i \omega_n )=
2 \gamma^2_{\bf k}\biggl[
(F_{\bf k}+\mu_B H)B( {\bf k} , i \omega_n )-
\delta_{\bf k}\Gamma( {\bf k} , i \omega_n )\biggr]
\eqno(16)
$$

We look upon the system of coupled
equations (9-11) as equivalent to the 
BCS gap equation in superconductivity. The 
kernels $V_{\bf k k'}$ in the different
CO, AFM or FM channels are input
parameters, like                  
the pairing potential is in BCS theory.
A solvable microscopic model
could in principle provide
the various kernels $V_{\bf k k'}$ for a given material 
system. Then we should solve a system of equations
(9-11) for each $\bf{Q}$ in the Brillouin zone.
The solution that minimizes the free energy 
characterized by a wave vector $\bf{Q}$ 
and a set of order parameters 
$W_{\bf k}$, $M_{\bf k}$ and $F_{\bf k}$,
will be the ground state of the system to be compared
with the experiments.
It results from the following analysis that
this is the 
correct procedure for the study
of the above orders in {\it any itinerant 
particle-hole asymmetric system} 
because in such systems the coexistence
and competition of these orders is shown     
to be unavoidable.         

Implementing the above procedure for a real system 
like manganites is a complex computational task
and requires a specific model asumption and 
perhaps related simplifications              
which are outside of the scope of this Letter.
Furthermore, for a detailled comparison with the
experiments 
in manganites additional elements like for example orbital
ordering may also be necessary to involve.
We focus in this Letter on stronger
qualitative arguments
which are valid whenever
itinerant FM, AFM and CO 
coexist 
whatever the exact microscopic physics is 
and are therefore valid for manganites as well.
With these arguments we explain only a
part of the physics of manganites
which 
results directly from
the competition of the above 
symmetry breakings.
However,
this part contains some
of the most fascinating puzzles in manganites 
including the CMR phenomenon.

We first note in (5) that
particle-hole symmetry                 
(i.e. $\delta_{\bf k}=0$)
implies $\Gamma_{\bf k}\propto W_{\bf k}M_{\bf k}$.
With this
one can show that 
if $\delta_{\bf k}=0$
then $W_{\bf k}=0$ is a trivial solution of (9),
$M_{\bf k}=0$ a trivial solution of (10) and
$F_{\bf k}=0$ a trivial solution of (11).
Therefore any combination of the above orders is possible.
Particle-hole
asymmetry induced by $\delta_{\bf k}\neq 0$
implies unexpected constraints.          
In fact,
let us start by considering
$F_{\bf k}=0=\mu_B H$. 
In both particle-hole
symmetric
($\delta_{\bf k}=0$) and particle-hole asymmetric
($\delta_{\bf k}\neq 0$) cases,
the trivial solutions 
$W_{\bf k}=0$ and $M_{\bf k}=0$
are independently valid in (9) and (10) respectively.
The situation is already 
different if we apply a uniform magnetic field
($\mu_B H\neq 0$).
For $\delta_{\bf k}=0$ the trivial
solutions $W_{\bf k}=0$ and $M_{\bf k}=0$ are still true
independently so that we may still have CO or AFM
alone at perfect nesting.
However, when we dope the system having $\delta_{\bf k}\neq 0$,
the trivial solutions $W_{\bf k}=0$ and $M_{\bf k}=0$
{\it are no more true independently}. We must either have both
$W_{\bf k},M_{\bf k}=0$ or both $W_{\bf k},M_{\bf k}\neq 0$ 
provided that none of $V_{\bf k k'}^{CO}$ and
$V_{\bf k k'}^{AFM}$ is identically zero which is the most natural
case for a real material system.
Applying a uniform magnetic field in a doped 
CO 
or AFM system 
we
{\it arrive at the 
coexistence of commensurate Charge and Spin Density orders}.

Let us
now take into account the possibility for FM ordering 
by considering also Eq. (11).
A similar analysis
can show that if $W_{\bf k}\neq 0$ and
$M_{\bf k}\neq 0$ and there is no particle-hole symmetry           
($\delta_{\bf k}\neq 0$),  
then $F_{\bf k}=0$ is {\it not} a trivial solution of (10).
Therefore $W_{\bf k},M_{\bf k}$ and $F_{\bf k}$ 
{\it necessarily
coexist} in a particle-hole asymmetric system. 
By doping the CO or AFM system we necessarily {\it
generate a ferromagnetic component}.
This may improve our understanding of 
FM in a variety of materials
ranging from transition metal compounds
like $MnSi$ \cite{Gilles} to borocarbides like
$TbNi_2B_2C$ \cite{FMboro}   
or organic materials like the doped fullerenes $TDAE-C_{60}$
(TDAE=tetrakis dimethyl amino ethylene) \cite{Lappas}
where signs of coexistence and/or competition
of FM with AFM-CO orders are evident.
Note that this result can 
be viewed as a formal generalization of the ``excitonic'' FM             
picture,
invoked recently 
for lightly
doped hexaborides \cite{lowDFM}.
We therefore stress the remarkable perspective that doped
hexaborides and manganites may have some
similar underlying physics.

We focus now on the behavior of the perovskite      
manganites.
In $La_{1-x}Ca_xMnO_3$ for example   
particle-hole symmetry ($\delta_{\bf k}=0$)
corresponds to $x=0.5$.
The metallic FM state is in competition with the
insulating AFM-CO state. 
They both occupy a portion of 
the carriers at finite doping.
If the AFM-CO state is melt,
the carriers from the AFM-CO state will be liberated and
the resistivity will drop.
The AFM-CO state
will be melt when
one of our quasiparticle poles
given in Eqs. (7) and (8) will go to zero,
in analogy with the estimate of the critical 
in-plane fields for the
melting of superconductivity          
in films \cite{Fulde}. 
When $W_{\bf k}\approx M_{\bf k}$, small
magnetic fields are sufficient
to melt the AFM-CO state even if $W_{\bf k}$ and $M_{\bf k}$
are very large. 
In fact, CO and AFM interfere producing 
quasiparticle poles with  
$W_{\bf k}+M_{\bf k}$ and $W_{\bf k}-M_{\bf k}$ the later
being the relevant since
these are likely to become zero.
We therefore consider the $W_{\bf k}-M_{\bf k}$ terms in 
(7) and (8), namely $E_{+-}({\bf k})$ and $E_{--}({\bf k})$.
We distinguish here two cases: Hole doping
corresponding to $\delta_{\bf k} < 0$  and electron doping for       
$\delta_{\bf k} >0$. In the case of hole doping, in
$E_{+-} ({\bf k})=\sqrt{\gamma_{\bf k}^2+(W_{\bf k}-M_{\bf k})^2}+
\delta_{\bf k}-F_{\bf k}-\mu_B H$, the doping $\delta_{\bf k}$,
the FM order $F_{\bf k}$ and the magnetic field $\mu_B H$ all have 
the same negative sign and cooperatively compete 
with the AFM-CO order making probable the  softening of 
the $E_{+-}({\bf k})$ branch and therefore the melting                 
of the AFM-CO order.
On the other hand, in the case of electron doping 
($\delta_{\bf k}>0)$,
in both relevant quasiparticle branches 
$E_{+-}({\bf k})$ and $E_{--}({\bf k})$, 
$\delta_{\bf k}$ will 
necessarily have {\it its sign oposite} to $F_{\bf k}$ and $\mu_B H$.
Therefore electron doping
does not cooperate with FM against the AFM-CO state
but instead {\it electron doping contributes to prevent the melting
of the AFM-CO order}.
This explains
the systematic difference
between electron-doped and hole-doped perovskite
manganites.            

Similarly, we can account  
for some basic aspects of the
recent experimental picture of perovskite manganites
drawn by
Roy, Mitchell, Ramirez and Schiffer \cite{x05f}.
Near half filling AFM-CO states 
and FM coexist since even
a small non-zero value of $\delta_{\bf k}$ generates a FM
component. Near half filling ($x\approx 0.5$) 
the dominating AFM-CO state can 
be melt by a small magnetic field because {\it the
critical temperatures of CO and AFM ordering 
coincide in the phase diagram of
perovskite manganites} (see 
for example Fig. 2 in \cite{millis})
indicating that indeed $W_{\bf k}$ and  $M_{\bf k}$
have similar magnitude and $E_{+-}$ may easily soften.
When we go with doping from the
hole doped regime to the slightly electron doped regime,
the critical field for the melting of the AFM-CO order 
increases because electron doping acts against the melting.
A marginal FM component exists even in the 
electron doped insulating state
and the melting of the AFM-CO state indeed
proceeds through an increase of the number of free carriers.

The 
CMR phenomenon can be understood
considering the 
relevant pole $E_{+-} ({\bf k})$.
At high hole doping, the FM order parameter $F_{\bf k}$ 
can be sufficiently large so that as it develops 
by lowering the temperature,
at $T=T_C$,
the pole
$E_{+-} ({\bf k})$ softens and the AFM-CO order is
melt liberating its portion of carriers and leading to the large 
enhancement of the conductivity.
The application of a magnetic field enhances the FM order
parameter from
$F_{\bf k}$ to $F_{\bf k}+\mu_B H$ and correspondingly
the melting critical temperature from $T_C$ to $T_C+\delta T_C$
producing 
negative CMR in the
temperature range $T_C< T < T_C+\delta T_C$.
In the above picture CMR is due to the
{\it  
increase of the number of carriers and not to a 
decrease in the scattering} in agreement
with the experiments \cite{x05f}.
The occurence of CMR in both perovskite and pyrochlore 
manganites demonstrates that CMR is not particular to a specific 
microscopic mechanism, which   
corroborates the above derived picture.

In summary, based on symmetry arguments 
associated with the coexistence
and competition of CO, AFM and FM orders              
we 
explain the particle-hole asymmetry in the phase diagram
of perovskite manganites and associate the melting of the AFM-CO
order and CMR to the similarity of CO and AFM  
order parameters.
Because of its generic (i.e. material independent) character
our analysis may help
the search for new magnetoresistive materials.
Our arguments should be valid
whenever these orders have
some itinerant character.                         
Therefore itinerant FM should normally be 
analyzed in the context of coexistence and competition with
AFM and CO orders as above.
This may improve our understanding of 
itinerant FM
in many different materials, and that of various related
problems like for example the melting of the spiral phases
in $MnSi$ \cite{Gilles} or the metamagnetic transitions
in $Y(Co_{1-x}Al_x)_2$ compounds \cite{Sakakibara}.

I am grateful to P. Fulde,
P.B. Littlewood and G. Lonzarich
for
stimulating and encouraging discussions, 
and a critical reading of the original manuscript.
I also acknowledge valuable discussions with T. Dahm, C. Geibel,
M. Lang, P.M. Oppeneer, A. Ovchinnikov,
M. Peter, G.C. Psaltakis, F. Steglich,
P. Thalmeier and G. Uimin.

\vskip 2cm



\begin{thebibliography}{999}

\bibitem{Ramirez} For a review see: A.P. Ramirez, Journal of Phys. C
{\bf 9}, 8171 (1997).

\bibitem{Jin} R. von Helmolt {\it et al.}
Phys. Rev. Lett. {\bf 71}, 2331 (1993);
S. Jin {\it et al.}, Science {\bf 264}, 413 (1994).

\bibitem{DE} C. Zener, Phys. Rev. {\bf 81},
440 (1951); P.W. Anderson and H. Hasegawa, Phys. Rev. {\bf 100},
675 (1955); P.-G. DeGennes, Phys. Rev. {\bf 118}, 141 (1960);
K.I. Kugel and D.I. Khomskii, JETP Lett. {\bf 15}, 445 (1972).

\bibitem{DE1} C.M. Varma, Phys. Rev. B {\bf 54}, 7328 (1996).

\bibitem{MLS} A.J. Millis, P.B. Littlewood and B. Shraiman,
Phys. Rev. Lett. {\bf 74}, 5144 (1995); H. Roder, J. Zhang and
A.R. Bishop, Phys. Rev. Lett. {\bf 76}, 1356 (1996);
A.J. Millis, R. Mueller and B.I. Shraiman, Phys. Rev. B {\bf 54},
5405 (1996).

\bibitem{millis} A.J. Millis,
Nature {\bf 392}, 147 (1998).

\bibitem{pyroexp} Y. Shimikawa, Y. Kubo and T. Manako,
Nature {\bf 379}, 53 (1996); S.-W. Cheong {\it et al.}, Solid State Commun.
{\bf 98} 163 (1996); M.A. Subramanian {\it et al.}, Science {\bf 273},
81 (1996).

\bibitem{noDE} A.P. Ramirez and M.A. Subramanian,
Science {\bf 277}, 546 (1997).

\bibitem{PBL} P. Majumdar and P.B. Littlewood, Phys. Rev. Lett.
{\bf 81}, 1314 (1998); Nature {\bf 395}, 479 (1998).

\bibitem{x05a} G. Papavassiliou {\it et al.}, Phys. Rev. B {\bf 55},
15000 (1997).

\bibitem{x05b} H. Kawano {\it et al.}, Phys. Rev. Lett. {\bf 78},
4253 (1997).

\bibitem{x05c} S. Mori, C.H. Chen and S.-W. Cheong,
Nature {\bf 392}, 473 (1998);
Phys. Rev. Lett. {\bf 81}, 3972 (1998).

\bibitem{x05d} P. Calvani {\it et al.}, 
Phys. Rev. Lett. {\bf 81}, 4504 (1998).

\bibitem{x05e} G. Allodi {\it et al.}, Phys. Rev. Lett. {\bf 81},
4736 (1998).

\bibitem{x05g} G. Kallias {\it et al.}, Phys. Rev. B {\bf 59},
1272 (1999). 

\bibitem{x05f} M. Roy {\it et al.}, Phys. Rev. B {\bf 58},
5185 (1998); J. Phys.: Cond. Matter {\bf 11}, 4843 (1999).

\bibitem{Liaro} E. Liarokapis {\it et al.}, 
Phys. Rev. B {\bf 60}, 12758 (1999).

\bibitem{Fath} M. F\"ath {\it et al.}, Science {\bf 285}, 1540 (1999).

\bibitem{Balagurov} A.M. Balagurov {\it et al.},
Phys. Rev. B {\bf 60}, 383 (1999).

\bibitem{Dagotto} A. Moreo, S. Yunoki and E. Dagotto,
Science {\bf 283}, 2034 (1999).

\bibitem{multi} 
P. Fulde and K. Maki, Phys. Rev. {\bf 141},
275 (1966); E.W. Fenton and G.C. Psaltakis,
Solid State Comm. {\bf 45}, 5 (1982).

\bibitem{Gilles} C. Pfleiderer {\it et al.}, Phys. Rev. B {\bf 55},
8330 (1997).

\bibitem{FMboro} B.K. Cho, P.C. Canfield and D.C. Johnston,
Phys. Rev. B {\bf 53}, 8499 (1996);
P. Dervenagas {\it et al.}, Phys. Rev. B {\bf 53}, 8506 (1996).

\bibitem{Lappas} P-M. Allemand {\it et al.}, Science
{\bf 253}, 301 (1991); A. Lappas {\it et al.}, Science {\bf 267},
1799 (1995).

\bibitem{lowDFM} Weak ferromagnetism in
$La$-doped hexaborides $CaB_6$, $SrB_6$ and $BaB_6$
(D.P. Young {\it et al.}, Nature {\bf 397},
412 (1999)) is recently discussed in the context of
a doped excitonic insulator.
(M.E. Zhitomirsky, T.M. Rice and
V.I. Anisimov, Nature {\bf 402}, 251 (1999);
L. Balents and C.M. Varma, Phys. Rev. Lett. {\bf 84},
1264 (2000);
V. Barzykin and L.P. Gorkov,
Phys. Rev. Lett. {\bf 84}, 2207 (2000)).
The possibility of ``excitonic'' FM has
first been reported by Volkov and Kopaev
(JETP Lett. {\bf 19}, 104 (1973))
in a two-band model for the excitonic insulator.

\bibitem{Fulde} P. Fulde, Adv. in Phys. {\bf 22},
667 (1973). 

\bibitem{Sakakibara} T. Sakakibara {\it et al.}, J. Phys.: Cond. Mat.
{\bf 2}, 3381 (1990).

\end{thebibliography}
\end{document}